\documentstyle[multicol,aps,psfig]{revtex}
\renewcommand{\narrowtext}{\begin{multicols}{2}
\global\columnwidth20.5pc\noindent}
\renewcommand{\widetext}{\end{multicols}
\global\columnwidth42.5pc}
\begin{document}
\draft
\preprint{\today}
\title{
Coexistent States of Charge Density Wave and 
Spin Density Wave 
in One-Dimensional Systems with 
the Inter-site Coulomb 
Interaction 
under the Electron Filling Control
}

\author{Keita Kishigi}
\address
{Faculty of Science, Himeji Institute of Technology,
 Ako, Hyogo 678-1297, Japan}
\date{\today}
\maketitle
\begin{abstract}
The coexistent state of the spin density wave (SDW) and
the charge density wave (CDW) in the one-dimensional 
systems 
is studied by the mean field approximation at $T=0$ in various 
electron-filling cases. 
We find that 
the coexistent state of 
SDW and CDW is stabilized when the on-site and the inter-site 
Coulomb interactions have the values estimated for the 
organic conductors. The ground state energies have cusp-like 
minima at 
1/4, 3/8, 5/12, 7/16, 7/20 and 9/20-fillings.

\end{abstract}
\narrowtext

\section{Introduction}
(TMTSF)$_2$$X$ and (TMTTF)$_2$$X$
($X$=ClO$_4$, PF$_6$, AsF$_6$, ReO$_4$, Br, SCN, etc.) 
have 
the quasi-one dimensional quarter-filled band, where 
superconductivity and the coexistent phase 
of charge density wave (CDW) and 
spin density wave (SDW) 
are observed.\cite{review,jerome}
In (TMTSF)$_2$PF$_6$ 
at the ambient pressure,
Pouget and Ravy observed 
the coexistence of $2k_{\rm F}$-SDW and
$2k_{\rm F}$-CDW 
by the X-ray scattering measurement,\cite{pouget} 
where $k_{\rm F}=\pi /4a$ 
is Fermi wave number and
$a$ is the lattice constant. 
Under pressure (12 kbar), the 
superconducting phase appears at 0.9 K.\cite{jerome2} 
In (TMTTF)$_2$Br, $4k_{\rm F}$-CDW accompanied by $2k_{\rm F}$-SDW
is found in X-ray scattering measurements.\cite{pouget} 
Moreover, the superconductivity transition occurs at 0.8 K 
under 26 kbar.\cite{3.62} 

These two kinds of the 
coexistent states ($2k_{\rm F}$-SDW-$2k_{\rm F}$-CDW and 
$2k_{\rm F}$-SDW-$4k_{\rm F}$-CDW) 
have been studied theoretically 
\cite{seofukuyama,nobuko,nobuko2,Mazumdar,yoshioka,tomio,tomio2,kishigi,kishigi2}. 
In these studies, 
they considered the 
inter-site Coulomb interaction ($V$) in addition to 
the on-site Coulomb interaction ($U$). 
Mila\cite{mila} has estimated $U/t\sim5$ and $V/t\sim2$, 
where 
$t$ is a transfer integral. 
The CDW-SDW coexistent state 
is understood as being caused by 
the interplay between $U$ and $V$.
\cite{seofukuyama,nobuko,nobuko2,Mazumdar,yoshioka,tomio,tomio2,kishigi,kishigi2,M.Nakamura} 
Seo and Fukuyama\cite{seofukuyama} showed that 
the ground state becomes the coexistent state of
$2k_{\rm F}$-SDW and $4k_{\rm F}$-CDW in the one-dimensional
extended Hubbard model with $V$. 
It is found by Kobayashi et al.\cite{nobuko,nobuko2} and 
Mazumdar et al.\cite{Mazumdar} that 
the coexistent state of $2k_{\rm F}$-SDW and $2k_{\rm F}$-CDW
is stabilized 
when the next nearest neighbor Coulomb interaction ($V_2$)
and the dimerization of the
energy band are considered. 
On the other hand, even if the 
dimerization is absent, it is indicated that 
the $2k_{\rm F}$-SDW and $2k_{\rm F}$-CDW 
coexists by Tomio and Suzumura\cite{tomio,tomio2} 
and the present author\cite{kishigi,kishigi2}.


The coexistent state of CDW and SDW 
is also found in quasi-two-dimensional
organic conductors such as (BEDT-TTF)$_2$$X$ 
which has the quarter filled band.\cite{q2d,q2d2,q2d3,q2d4}
In quasi-two-dimensional organic conductors, 
the coexistent state of CDW and SDW is observed as 
the stripe-type order. 
The CDW-SDW coexistent state in quasi-two-dimensional systems 
is theoretically studied by Seo.\cite{seo}
Recently, it is found that the ground state in $X$=KHg changes from 
the charge order\cite{miyagawa16}, which makes 
the stripe-type order, 
to the superconductivity 
under uniaxial pressure.\cite{maesato} 
Since, even in 
quasi-one-dimensional quarter-filled systems such as 
(TMTSF)$_2$$X$ and (TMTTF)$_2$$X$, 
the coexistent state of CDW and SDW 
changes to the superconductivity 
under pressure,\cite{jerome2,3.62} 
it is expected that 
the superconductivity appears near the coexistent state of 
CDW and SDW. 
In the 
high-$T_c$ cuprates such as La$_{1.6-x}$Nd$_{0.4}$Sr$_x$CuO$_4$,\cite{nature} 
the stripe-type order 
is found when the rate of the doping is $x=1/8$ 
and the critical temperature of the superconductivity becomes the highest 
at $x\simeq 1/8$. 
The superconductivity near the stripe-type order 
may be 
given by the fluctuation of 
the antiferromagnetism of the stripe-type order, because 
it is understood that the superconductivity 
in high-$T_c$ cuprates near the half-filling at which the 
the antiferromagnetic state appears 
is due to 
the strong fluctuation of the antiferromagnetism.\cite{anderson}




In this paper, we study how the elelctron-filling ($f$) 
affects 
the coexistent state of SDW and CDW due to $V$ 
in the mean field theory. 
By calculating the condensation energy of the 
coexistent state, 
we find $f$ at which the coexistent state 
in one-dimensional systems is stabilized. 
The $f$-dependence of the condensation energy
in the one-dimensional extended Hubbard model 
has never been studied, 
although the ground state energy as a function of $f$ 
has been calculated in the one-dimensional 
Hubbard model.\cite{carmelo}
Based on the result, we propose $f$ which favors the 
superconductivity in the strongly 
correlated one-dimensional systems with $V$.

We use the one-dimensional extended Hubbard model
with $V$, where 
the dimerization is neglected for simplicity, 
because the coexistent states are stabilized without 
the dimerization.\cite{tomio,tomio2,kishigi,kishigi2} 
We use parameter, $U/t=5.0$, 
\cite{band,band2,band3,band4} 
and vary the parameter $V$ in the region of 
$0\leq V\leq U$.




\section{Model}

The one-dimensional extended Hubbard model is, 
\begin{eqnarray}
\hat{\cal H}&=&\hat{\cal K}+\hat{\cal U}+\hat{\cal V}, \\
\hat{\cal K}&=&-t\sum_{i,\sigma}(c^{\dagger}_{i,\sigma} c_{i+1,\sigma}+
h.c.), \\
\hat{\cal U}&=&U\sum_{i}n_{i, \uparrow}n_{i, \downarrow}, \\
\hat{\cal
V}&=&V\sum_{i,\sigma,\sigma^{\prime}}n_{i,\sigma}n_{i+1,\sigma^{\prime
}}, 
\end{eqnarray}
where 
$c^{\dagger}_{i,\sigma}$ is the creation
operator of $\sigma$ spin electron at $i$ site,
$n_{i,\sigma}$ is the number operator, 
$i=1,\cdots,N_{\rm S}$, $N_{\rm S}$ is the
number of the total sites and $\sigma =\uparrow$ and $\downarrow$.

The interaction terms, $\hat{\cal U}$ and 
$\hat{\cal V}$, are treated in the mean field
approximation as 
\begin{eqnarray}
\hat{\cal U}^{\rm M}&=&\sum_{k_{x}}
\sum_{Q}\{
\rho_{\uparrow\uparrow}(Q)
C^{\dagger}(k_{x},
\downarrow)
C(k_{x}-Q,\downarrow) \nonumber \\ 
&+&\rho^{*}_{\downarrow\downarrow}(Q)
C^{\dagger}(k_{x}-Q,
\uparrow)
C(k_{x},\uparrow)\}, \\
\hat{\cal V}^{\rm M}&=&(\frac{V}{U})\sum_{k_{x},
\sigma,\sigma^{\prime}}\sum_{Q}e^{-iQa}\{
\rho_{\sigma\sigma}(Q)
C^{\dagger}(k_{x},
\sigma^{\prime})
C(k_{x}-Q,
\sigma^{\prime}) \nonumber \\
&+&\rho_{\sigma^{\prime}\sigma^{\prime}}^{*}(Q)
C^{\dagger}(k_{x},
\sigma)
C(k_{x}-Q,
\sigma)\}
\end{eqnarray}
, where $I=U/N_{\rm S}$.
The self-consistent equation for the order parameter 
$\rho_{\sigma\sigma}(Q)$ 
is given by
\begin{eqnarray}
\rho_{\sigma\sigma}(Q)=I\sum_{k_{x}}
<C^{\dagger}(k_{x},
\sigma)
C(k_{x}-Q,
\sigma)>.
\end{eqnarray}
We do not consider the mean field, $\rho_{\sigma\bar{\sigma}}(Q)$, 
where $\bar{\sigma}$ is the opposite 
spin of ${\sigma}$ 
for the simplicity.
We consider the various electron-fillings, $f=q/p$, 
where $q$ and $p$ 
are mutually prime numbers. 
We take the possible wave vectors of the 
order parameters as 
$Q=nQ_0$, where 
$Q_0=2k_{\rm F}=(2\pi /a)\cdot f$ and $n=1,\cdots,p$. 
The matrix size of the mean field Hamiltonian, 
$\hat{\cal K}+\hat{\cal U}^{\rm M}+\hat{\cal V}^{\rm M}$, 
is $p\times p$. 
We calculate the self-consistent solutions by 
using eigenvectors obtained by diagonalizing
$\hat{\cal K}+\hat{\cal U}^{\rm M}+\hat{\cal V}^{\rm M}$.


The ground state energy per site is 
\begin{eqnarray}
E_g(\rho_{\sigma\sigma}(Q))&=&\frac{1}{N_s}\sum^{N_{s}/2}_{j=1}\epsilon_{j}
-\frac{1}{U}\sum_{Q}\rho_{\uparrow\uparrow}(Q,0)\rho^{*}_
{\downarrow\downarrow}(Q,0) \nonumber \\ 
&-&\frac{V}{U^2}\sum_{Q,\sigma,\sigma^{\prime}}e^{-iQa}\rho_{\sigma\sigma}(Q,0)
\rho_{\sigma^{\prime}\sigma^{\prime}}^{*}(Q,0), 
\end{eqnarray}
where $\epsilon_{j}$ is the eigenvalue and 
the index $j$ includes the degree of the spin freedom.
The condensation energy, $E_c$, is given by 
$E_c=E_g-E_N$, where $E_N$ is the normal state energy.

The electron density ($n(j)$) 
and the 
spin moment ($S_z(j)$) at site $j$ are given by
\begin{eqnarray}
n(j)=\frac{1}{U}\sum_{Q,\sigma}\rho_{\sigma\sigma}(Q)e^{iQja}
\end{eqnarray}
and 
\begin{eqnarray}
S_z(j)=\frac{1}{2U}\sum_{Q}(\rho_{\uparrow\uparrow}(Q)
-\rho_{\downarrow\downarrow}(Q))e^{iQja}.
\end{eqnarray}



\begin{figure}
\mbox{\psfig{figure=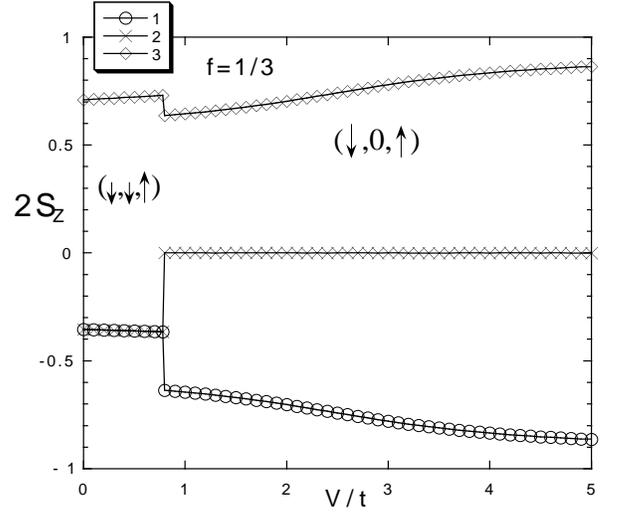,width=80mm,angle=0}}
\vskip 5mm
\caption{At $f=1/3$, 2S$_z$(1), 2S$_z$(2) and 2S$_z$(3) as a function of $V/t$.
}
\label{fig1a}
\end{figure}
\vskip 3mm

\begin{figure}
\mbox{\psfig{figure=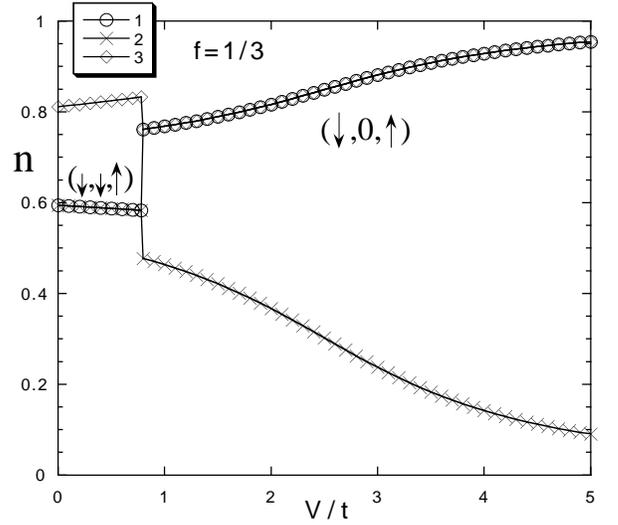,width=80mm,angle=0}}
\vskip 5mm
\caption{At $f=1/3$, $n(1)$, $n(2)$ and $n(3)$ as a function of $V/t$. 
}
\label{fig1b}
\end{figure}
\vskip 3mm

\section{Results and Discussions}

We search the most stable self-consistent solutions by 
changing the initial values of order parameters. 
Since $n(j)$, $S_z(j)$ and $E_c$ for $f>0.5$ is the same as those 
for $f<0.5$ due to the symmetry between 
an electron and an hole and we do not 
focus our attention on the low electron-filling, we calculate 
at various fillings ($2\leq p\leq 20, 1\leq q\leq9$) 
in the region of $0.2\leq f\leq0.5$. 

At $f=1/2$, 
the stable state is 2$k_{\rm F}$-SDW, ($\downarrow,\uparrow)$, 
($S_z(1)=-S_z(2)\simeq1.0$ and $n(1)=n(2)\simeq1.0$) 
for $2V\leq U$ and the state is changed to 
2$k_{\rm F}$-CDW, (0,$\downarrow\uparrow)$, 
($S_z(1)=S_z(2)\simeq0$ and $n(1)\simeq0$ and $n(2)\simeq2.0$) 
for $2V>U$, where 
the arrows mean the spin moment, 0 means small or zero 
electron density 
and $\downarrow \uparrow$ represents 
that the up and down electrons exist in the same site. 
For $2V>U$, 
the period of 
the CDW becomes two. 
As this charge order is suitable for the nearest Coulomb repulsion interaction, 
the large charge gap made by the CDW exists at large $V$. 
As a result, $E_c$ becomes lower, which will be 
shown in Fig. 7. 
This result at the half-filling 
has been shown by Cabib and Callen.\cite{1/2}.

At $f=1/3$, the stable state is the coexistence of SDW and 
CDW, $(_{\downarrow},_{\downarrow},\uparrow)$, 
($|S_z(1)|=|S_z(2)|<S_z(3)$ and $n(1)=n(2)<n(3)$) for $0\leq V/t< 0.8$, where 
the distortion of the charge density is small and 
the coexistent state of SDW and CDW,
$({\downarrow},0,\uparrow)$ 
$(S_z(1)=-S_z(3), S_z(2)=0$, $n(1)=n(3)\simeq 1.0$ and $n(2)\simeq 0$), 
is stabilized for $V/t\geq 0.8$, as shown in Figs. 1 and 2. 
As the period of $n$ in $({\downarrow},0,\uparrow)$ is 
three, there is a frustration for $V$ and 
the energy gap dose not become large. 
Since the order parameters having $Q=(2\pi /a)\cdot(1/2)$ 
dose not exist when $p$ of $f=q/p$ is odd 
number, the period of $n$ dose not become two. 
Thus, when $p$ is odd number, 
the order of the CDW is not suitable for 
$V$.

\begin{figure}
\mbox{\psfig{figure=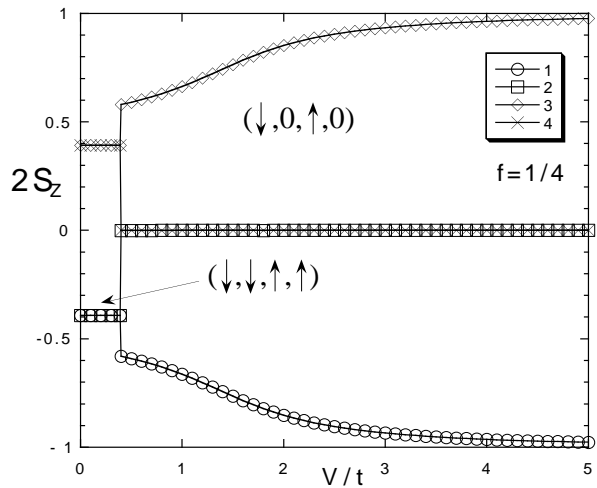,width=80mm,angle=0}}
\vskip 5mm
\caption{At $f=1/4$, 2S$_z(j)$ $(j=1\cdots 4)$ as a function of $V/t$. 
}
\label{fig1a}
\end{figure}
\vskip 3mm

\begin{figure}
\mbox{\psfig{figure=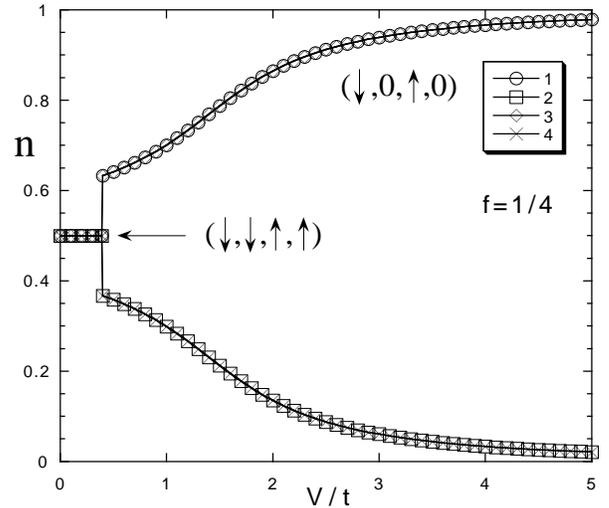,width=80mm,angle=0}}
\vskip 5mm
\caption{At $f=1/4$, $n(j)$ $(j=1\cdots 4)$ as a function of $V/t$.
}
\label{fig1b}
\end{figure}
\vskip 3mm






In the case of the quarter filling ($f=1/4$), 
the stable state of 2$k_{\rm F}$-SDW, 
$({\downarrow},{\downarrow},\uparrow,\uparrow)$ 
$(S_z(1)=S_z(2)=-S_z(3)=-S_z(4)$ and $n(1)=n(2)=n(3)=n(4)=0.5$), 
is changed to the coexistent state of 2$k_{\rm F}$-SDW and 
4$k_{\rm F}$-CDW, $({\downarrow},0,\uparrow,0)$ 
($S_z(1)=-S_z(3), S_z(2)=S_z(4)=0$, 
$n(1)=n(3)\simeq 1.0$ and $n(2)=n(4)\simeq 0$), for $V/t>0.39$, 
as shown in Figs. 3 and 4. 
This result is consistent with the previous study 
by Seo and Fukuyama.\cite{seofukuyama} 
For $V/t>0.39$, as the 
period of the charge density is two, 
$E_c$ becomes large 
due to the large energy gap, which will be shown in Fig. 7.

We show 2$S_z$ and $n$ at $f=3/8$ in Figs. 5 and 6, where 
$S_z(1)=-S_z(5)$, $S_z(2)=-S_z(6)$, $S_z(3)=-S_z(7)$, $S_z(4)=-S_z(8)$, 
$n(1)=n(5)$, $n(2)=n(6)$, $n(3)=n(7)$ and $n(4)=n(8)$. 
The stable coexistent state of SDW and CDW is 
$(_{\uparrow},_{\uparrow},{\downarrow},{\uparrow},_{\downarrow},_{\downarrow},{\uparrow},{\downarrow})$ for $V/t<0.7$,
$(0,{\uparrow},{\downarrow},{\uparrow},0,{\downarrow},{\uparrow},{\downarrow})$ for $0.7\leq V/t\leq3.0$ and 
$(0,{\uparrow}_{\downarrow},0,{\uparrow}_{\downarrow},0,{\downarrow}_{\uparrow},0,{\downarrow}_{\uparrow})$ ($S_z(1)=S_z(3)=S_z(5)=S_z(7)=0$ and 
$S_z(2)=S_z(4)=-S_z(6)=-S_z(8)$) for $V/t>3.0$.
In particular, the period of $n$ 
($n(1)=n(3)=n(5)=n(7)\simeq 0$ and $n(2)=n(4)=n(6)=n(8)\simeq 1.5$) 
is two for $V/t>3.0$, 
which gives the large gain of $E_c$, 
as will be shown in Fig. 7. 

\begin{figure}
\mbox{\psfig{figure=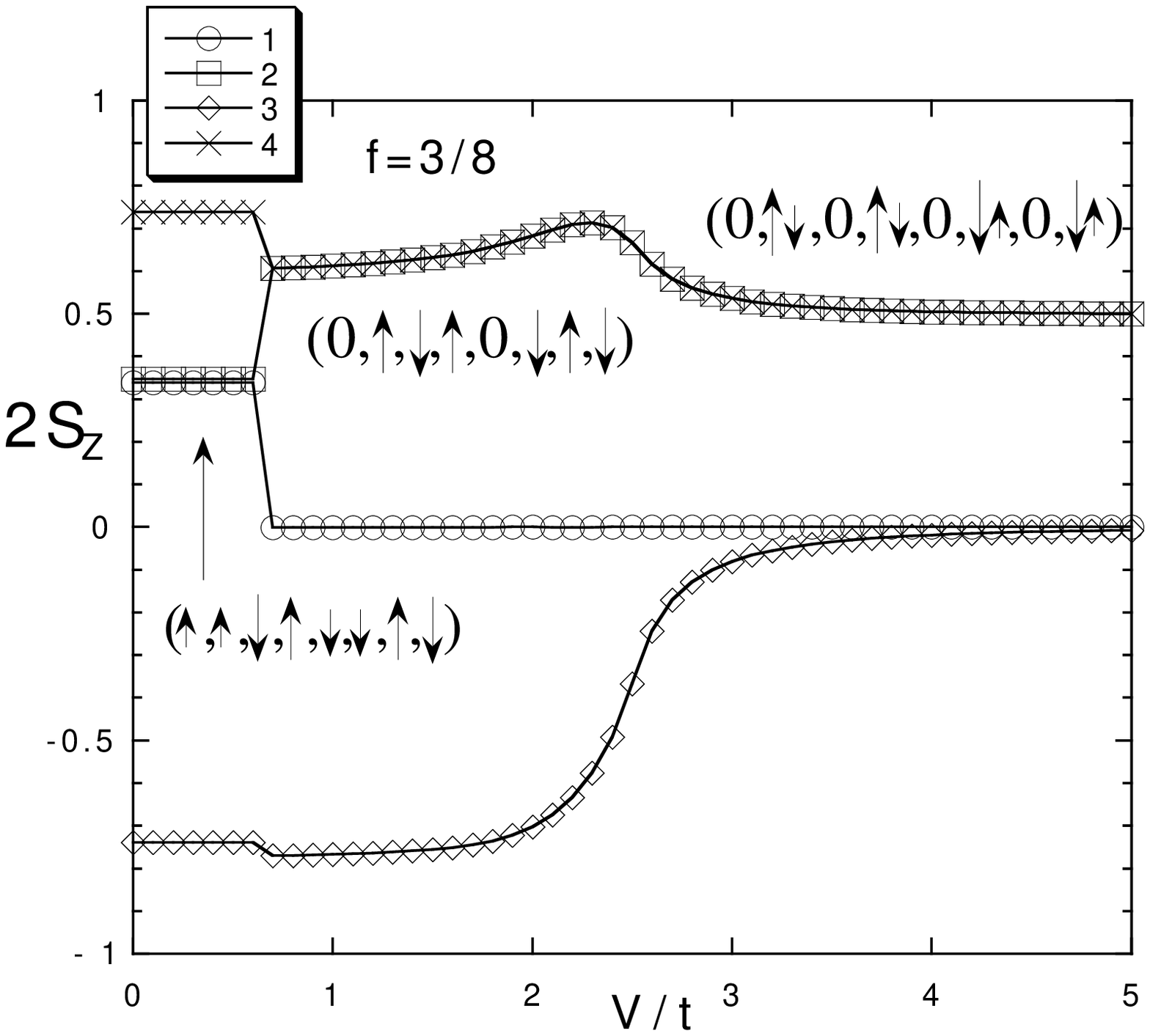,width=80mm,angle=0}}
\vskip 5mm
\caption{At $f=3/8$, 2S$_z(j)$ $(j=1,\cdots,4)$ as a function of $V/t$. 
}
\label{fig1b}
\end{figure}
\vskip 3mm

\begin{figure}
\mbox{\psfig{figure=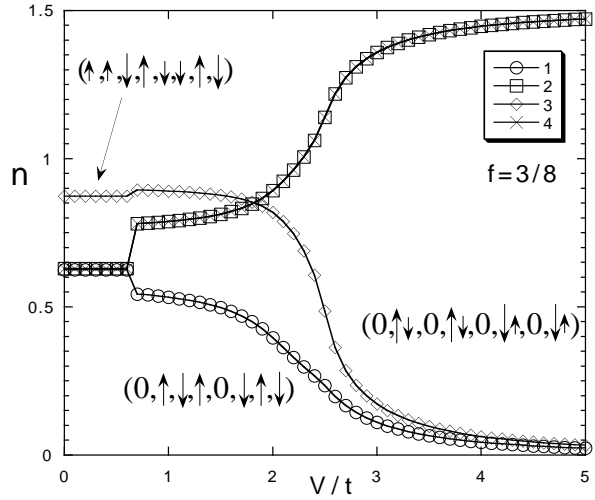,width=80mm,angle=0}}
\vskip 5mm
\caption{At $f=3/8$, $n(j)$ $(j=1,\cdots,4)$ as a function of $V/t$.
}
\label{fig1b}
\end{figure}
\vskip 3mm

In the case of other fillings except the half-filling, 
the coexistent state of CDW and SDW is stabilized.

Next, we show $E_c$ in Fig. 7 as a function of 
$f$ for various $V$. 
At $V=0$, $E_c$ monotonically decreases as $f$ increases, because 
$E_c$ is mainly determined by 
the density of states on 
the Fermi surface, $N(0)=N_{\rm S}/(4\pi t\sin ak_{\rm F})$, 
which decreases as $f$ increases. 
It is found that $E_c$ at $f=1/4$ rapidly decreases as $V$ 
increases. 
In this filling, the localization of electrons 
occurs easily even at weak $V$. 
The reason is that the order of $({\downarrow},0,\uparrow,0)$ is 
favorable for the 4$k_{\rm F}$-CDW 
keeping the 2$k_{\rm F}$-SDW state. 
We can see that $|E_c|$ for $f=1/2, 1/4, 3/8, 5/12$, 
7/16, 7/20 and 9/20 become large, 
compared to other fillings, upon increasing $V$. 
It is found that the cusp-like minima of the 
condensation energy 
in the coexistent state of CDW and SDW appear at 
$f=n/4m$, where 
$n$ and $m$ are integers. 
In these fillings, the period of $n$ 
accompanying the order of the SDW becomes 
two, resulting in 
the large energy gap. 
The local minimum at $f=5/16$ may be too small to see 
in Fig. 7. 

\begin{figure}
\mbox{\psfig{figure=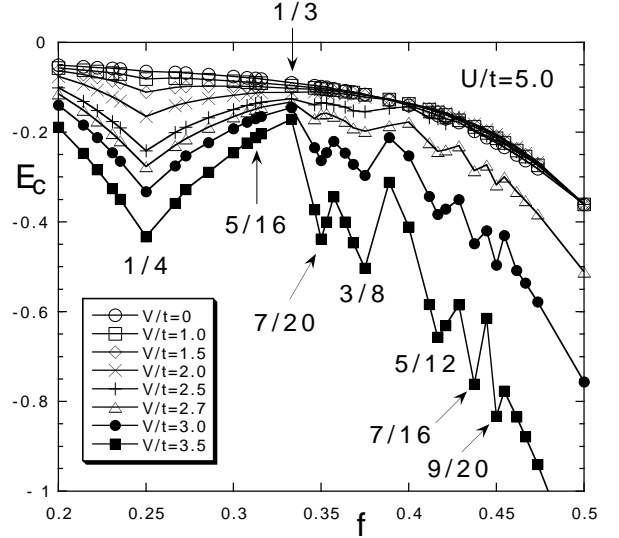,width=80mm,angle=0}}
\vskip 5mm
\caption{
$E_c$ as a function of 
$f$ for 
various $V$.
}
\label{fig3}
\end{figure}
\vskip 3mm

In the coexistent state of CDW and SDW, 
the commensurabity of the electron-filling and the inter-site Coulomb 
interaction play the important role of the 
condensation energy. 
At 
$f=n/4m$, 
the coexistent state of CDW and SDW is 
compatible with the favored real-space order for both $U$ and $V$. 
On the other hand, $E_c$ at $f=1/3$ changes little as $V$ increases. 
This feature can be understood by considering that the 
frustration occurs in the order of period three; 
double occupancy should be reduced by $U$ and electorns at 
nearest sites should be avoided for large $V$. 

In the case of one-dimensional systems with 
$V$ where 
the superconductivity is attributable to 
the fluctuation of 
the antiferromagnetism, 
the superconductivity may appear at the fillings near $f=n/4m$. 
If the origin of 
the superconductivity of quasi-one-dimensional organic conductors such as 
(TMTTF)$_2$$X$ and (TMTSF)$_2$$X$ is so, 
the superconductivity 
will be obtained 
by changing electron-filling slightly from the quarter-filling.


\section{Conclusion}
We study 
the coexistent state of CDW and SDW in the 
one-dimensional system with 
the inter-site Coulomb interaction 
by changing the electron-filling. 
We find that 
the 
coexistent state is stabilized 
at $f=n/4m$, for example, 
1/4, 3/8, 5/12, 7/16, 7/20 and 9/20-fillings
due to the inter-site Coulomb interaction. 
Since the strong fluctuation of the 
antiferromagnetism is expected in the region close to 
the coexistent state, 
the superconductivity will appear near $f=n/4m$.



The author would like to thank
Y. Hasegawa 
for valuable discussions.

\widetext

\begin{references}
\bibitem{review}
For a review, see: T. Ishiguro, K. Yamaji, and G. Saito: {\it
Organic
Superconductors}
(Springer-Verlag, Berlin  1998).

\bibitem{jerome}
For a review, see D. Jerome: {\it
Organic
Conductors} ed J. P. Farges
(Marcel Deckker, New York, 1994).




\bibitem{pouget}
J. P. Pouget and S. Ravy:
Synth. Met. {\bf 85} (1997) 1523.

\bibitem{jerome2}
D. Jerome, 
A. Mazaud, M. Ribault, 
K. Bechgaard: 
J. Physique Lett. 
{\bf 41} (1980) L95.

\bibitem{3.62}
L. Balicas, 
K. Behnia, W. Kang, E. Canadell, 
P. Auban-Senzier, D. Jerome, M. Ribalut, 
J. M. Fabre: 
J. Phys. I (France) 
{\bf 4} (1994) 1539.



\bibitem{seofukuyama}
H.Seo and H. Fukuyama: J. Phys. Soc. Jpn. {\bf 66}
(1997) 1249.

\bibitem{nobuko}
N. Kobayashi and M. Ogata: J. Phys. Soc. Jpn. {\bf 66}
(1997) 3356.


\bibitem{nobuko2}
N. Kobayashi, M. Ogata and K. Yonemitsu:
J. Phys. Soc. Jpn.
{\bf 67} (1998) 1098.


\bibitem{Mazumdar}
S. Mazumdar, 
S. Rammasesha, R. Torsten Clay and
David K. Campbell:
Phys. Rev. Lett. {\bf 82} (1999) 1522.

\bibitem{yoshioka}
H. Yoshioka, M. Tsuchiizu, and Y. Suzumura, 
J. Phys. Soc. Jpn. {\bf 69} (2000) 651.


\bibitem{tomio}
Y. Tomio and Y. Suzumura,
J. Phys. Soc. Jpn.
{\bf 69} (2000) 796.

\bibitem{tomio2}
Y. Tomio and Y. Suzumura,
preprint.

\bibitem{kishigi}
K. Kishigi and Y. Hasegawa: J. Phys. Soc. Jpn.
{\bf 69} (2000) No. 7.

\bibitem{kishigi2}
K. Kishigi and Y. Hasegawa: cond-mat/0003502.

\bibitem{M.Nakamura}
For the recent calculation of the exact diagonalization, 
M. Nakamura: cond-mat/9909277

\bibitem{mila}
F. Mila:
Phys. Rev. B {\bf 52} (1995) 4788.



\bibitem{q2d}
C. P. Heidmann, 
A. Barnsteiner, F. Gro$\beta$-alltag, B. S. 
Chandrasekhar and E. Hess: 
Solid State Commun.
{\bf 84} (1992) 711.

\bibitem{q2d2}
J. Moldenhauer, 
Ch. Horn, K. I. Pokodinia, D. Schweitzer, 
I. Heiden and H. J. Kelle: 
Synth. Met. {\bf 60} (1993) 31.

\bibitem{q2d3}
K. Miyagawa, A. Kawamoto and K. Kanoda: preprint

\bibitem{q2d4}
T. Nakamura, 
W. Minagawa, R. Kinami and T. Takahashi: 
J. Phys. Soc. Jpn.
{\bf 69} (2000) 504.

\bibitem{seo}
H. Seo: J. Phys. Soc. Jpn. {\bf 69} (2000) 805.

\bibitem{miyagawa16}
K. Miyagawa, A. Kawamoto and K. Kanoda: 
Phys. Rev. B{\bf 56} (1997) 8487.

\bibitem{maesato}
M Maesato: Doctor Thesis in Tokyo-University in Japan (2000)

\bibitem{nature}
J. M. Tranquada et al.: Nature {\bf 375} (1995) 561. 

\bibitem{anderson}
P. W. Anderson: 
{\it The Theory of Superconductivity in the High-T$_c$ Cuprates
}
(Princeton University Press, NY, 1996).









\bibitem{carmelo}
J. Carmelo and D. Baeriswyl: Phys. Rev. B{\bf 37} (1988) 7541.


\bibitem{band}
T. Mori, 
A. Kobayashi, Y. Sasaki and H. Kobayashi:
Chem. Lett. (1982) 1923.

\bibitem{band2}
P. M. Grant: J. Phys. Colloq. C{\bf 3} (1983) 847.

\bibitem{band3}
T. Mori, 
A. Kobayashi, Y. Sasaki and H. Kobayashi, G. Saito and
H. Inokuchi: Bull.
Chem. Soc. Jpn. {\bf 57} (1984) 627.

\bibitem{band4}
L. Ducasse, 
M. Abderrabba, J. Hoarau, M. Pesquer,
B. Gallois and J. Gaultier: 
J. Phys. C{\bf 19} (1986) 3805.

\bibitem{1/2}
D. Cabib and E. Callen: Phys. Rev B 12 (1975) 5249.












\end{references}
\end{document}